\documentclass[runningheads]{llncs}

\usepackage{graphicx}
\usepackage{amsmath}
\usepackage{xspace}
\usepackage{booktabs}
\usepackage{url}
\usepackage{amsfonts}
\usepackage{amssymb}
\usepackage{multirow}
\usepackage[ruled,vlined]{algorithm2e}
\usepackage{enumitem}
\usepackage{numprint}
\usepackage{tcolorbox}
\usepackage{xurl}
\usepackage{xstring}
\usepackage[T1]{fontenc}
\usepackage{pgfplots}

\usepackage{color}

\usepackage[colorlinks,citecolor=.]{hyperref}

\newcommand{\liquidator}{liquidator\xspace}
\newcommand{\liquidators}{liquidators\xspace}

\newcommand{\borrower}{borrower\xspace}

\newcommand{\lender}{lender\xspace}

\newcommand{\supporter}{supporter\xspace}
\newcommand{\supporters}{supporters\xspace}
\newcommand{\Supporter}{Supporter\xspace}
\newcommand{\Supporters}{Supporters\xspace}

\newcommand{\seller}{seller\xspace}

\newcommand{\sellerShort}{\mathcal{C}_S}

\newcommand{\buyer}{buyer\xspace}

\newcommand{\buyerShort}{\mathcal{C}_B}

\definecolor{applegreen}{rgb}{0.55, 0.71, 0.0}
\definecolor{bleudefrance}{rgb}{0.19, 0.55, 0.91}
\newcommand{\colorEntity}[1]{{\color{brown}{#1}}}
\newcommand{\colorEntityBorrower}[1]{{\color{applegreen}{#1}}}
\newcommand{\colorEntitySupporter}[1]{{\color{bleudefrance}{#1}}}
\newcommand{\liquidatorShort}{\colorEntity{\mathcal{Q}}}
\newcommand{\borrowerShort}{\colorEntityBorrower{\mathcal{B}}}
\newcommand{\lenderShort}{\colorEntity{\mathcal{E}}}
\newcommand{\supporterShort}{\colorEntitySupporter{\mathcal{S}}}

\newcommand{\financialPrimitive}{reversible call option\xspace}
\newcommand{\financialPrimitives}{reversible call options\xspace}
\newcommand{\FinancialPrimitive}{Reversible Call Option\xspace}

\newcommand{\lendingPool}{lending pool\xspace}

\newcommand{\lendingPoolShort}{\mathcal{L}}

\newcommand{\borrowingPositionShort}[1]{P_{#1}}
\newcommand{\borrowingPosition}{borrowing position\xspace}

\newcommand{\borrowingPositions}{borrowing positions\xspace}

\newcommand{\numberCoinsDebt}[1]{D_{#1}}

\newcommand{\numberCoinsCollateral}[1]{C_{#1}}

\newcommand{\price}{p_t}
\newcommand{\priceInp}[1]{p_{#1}}

\newcommand{\healthFactor}{health factor\xspace}

\newcommand{\healthFactorShort}[1]{HF_t(#1)}
\newcommand{\healthFactorShortTwo}[2]{HF_{#1}(#2)}
\newcommand{\collateralizationRatio}{collateralization ratio\xspace}

\newcommand{\collateralizationRatioShort}[1]{CR_t(#1)}
\newcommand{\collateralizationRatioShortTwo}[2]{CR_{#1}(#2)}
\newcommand{\collateralDiscount}{collateral discount\xspace}

\newcommand{\collateralDiscountShort}{\theta}

\newcommand{\closeFactor}{close factor\xspace}

\newcommand{\closeFactorShort}{k_{CF}\xspace}
\newcommand{\liquidationSpread}{liquidation spread\xspace}

\newcommand{\liquidationSpreadShort}{S}
\newcommand{\protocolShort}{\texttt{prot}}
\newcommand{\fixedSpreadProtocol}{\protocolShort_{FSL}}

\newcommand{\amountAssetShort}{N}
\newcommand{\AssetShort}{A}
\newcommand{\strikePriceShort}{K}
\newcommand{\premiumShort}{\phi}
\newcommand{\factorPrimitive}{k}
\newcommand{\terminationTime}{T}
\newcommand{\Time}{t}
\newcommand{\initiateTime}{t_0}
\newcommand{\factorMiqado}{\lambda}
\newcommand{\factorMiqadoRe}{k_{re}}
\newcommand{\returnPaymentShort}{\numberCoinsCollateral{re}}
\newcommand{\borrowingInterestRate}{I_{\lendingPoolShort}}

\newcommand{\mainIndent}{\noindent\hspace*{30pt}\ignorespaces}
\newcommand{\mainIndentTwo}{\noindent\hspace*{70pt}\ignorespaces}

\newcommand{\spotExchangeRate}{S_0}
\newcommand{\foreignInterestRate}{r_f}
\newcommand{\domesticInterestRate}{r}
\newcommand{\volatilityAsset}{\sigma}

\newcommand{\bufferShort}{B}
\newcommand{\supportFactorShort}{k_{SF}}

\newcommand{\liquidationSpiral}{liquidation spiral\xspace}
\newcommand{\liquidationSpirals}{liquidation spirals\xspace}
\newcommand{\LiquidationSpiral}{Liquidation Spiral\xspace}

\newcommand{\etal}{\textit{et al.}\xspace}
\npthousandsep{,}
\npdecimalsign{.}

\newcommand{\empirical}[1]{#1}

\newcommand{\etherscantx}[1]{\href{https://etherscan.io/tx/#1}{\wrapletters{#1}}\xspace}

\newcommand*\wrapletters[1]{\wr@pletters#1\@nil}
\def\wr@pletters#1#2\@nil{#1\allowbreak\if&#2&\else\wr@pletters#2\@nil\fi}

\newcommand{\protocol}{\textsc{Miqado}\xspace}

\newcommand{\defiMarketSize}{$50$B~USD\xspace}
\newcommand{\defiLendingMarketSize}{$15$B~USD\xspace}
\newcommand{\defiLendingProportion}{$30\%$\xspace}

\newcommand{\StartDate}{\empirical{1st of May,~2019}\xspace}
\newcommand{\EndDate}{\empirical{30th of September,~2022}\xspace}
\newcommand{\liquidationTimeFrame}{\empirical{$41$~months}\xspace}
\newcommand{\AaveVOneLiquidations}{\empirical{$\numprint{5765}$}\xspace}
\newcommand{\AaveVTwoLiquidations}{\empirical{$\numprint{25576}$}\xspace}
\newcommand{\CompoundLiquidations}{\empirical{$\numprint{17023}$}\xspace}
\newcommand{\TotalLiquidationEvents}{\empirical{$\numprint{48364}$}\xspace}

\newcommand{\LiquidatedCollateral}{\empirical{$\numprint{2.32}$B~USD}\xspace}
\newcommand{\CollateralReleasePeak}{\empirical{$\numprint{653.11}$M~USD}\xspace}
\newcommand{\ShortLiquidations}{\empirical{$\numprint{18305}$}\xspace}
\newcommand{\ShortLiquidationUSDSold}{\empirical{$1.33$B~USD}\xspace}
\newcommand{\FullySoldShortLiquidations}{\empirical{$\numprint{3365}$}\xspace}
\newcommand{\ShortLiquidationsSellPercentageAverage}{\empirical{$95.95\%$}\xspace}
\newcommand{\ShortLiquidationPriceDeclineAverage}{\empirical{$0.38\%$}\xspace}
\newcommand{\ShortLiquidationPriceDeclineMax}{\empirical{$26.90\%$}\xspace}

\newcommand{\CollateralRestraintTwenty}{\empirical{$5.63$B}\xspace}
\newcommand{\CollateralRestraintTen}{\empirical{$2.82$B}\xspace}
\newcommand{\CollateralRestraintFive}{\empirical{$1.41$B}\xspace}
\newcommand{\CollateralRestraintTwo}{\empirical{$563.40$M}\xspace}
\newcommand{\CollateralRestraintOne}{\empirical{$281.70$M}\xspace}

\newcommand{\HealthyPositionsAfterFSL}{\empirical{$82.25\%$}\xspace}
\newcommand{\HealthyPositionsAfterMiqadoLambdaFive}{\empirical{$82.22\%$}\xspace}

\newcommand{\CollateralReleaseWithMiqado}{\empirical{$236.40$M~USD}\xspace}
\newcommand{\CollateralReduction}{\empirical{$89.82\%$}\xspace}

\usepackage{acro}
\acsetup{single}

\DeclareAcronym{DeFi}{
  short = DeFi,
  long  = Decentralized Finance,
}
\newcommand{\DeFi}{\ac{DeFi}\xspace}

\DeclareAcronym{FSL}{
  short = FSL,
  long  = fixed spread liquidation,
}
\newcommand{\FSL}{\ac{FSL}\xspace}

\DeclareAcronym{TVL}{
  short = TVL,
  long  = total value locked,
}
\newcommand{\TVL}{\ac{TVL}\xspace}

\DeclareAcronym{MEV}{
  short = MEV,
  long  = miner extractable value,
}
\newcommand{\MEV}{\ac{MEV}\xspace}

\begin{document}

\title{Mitigating Decentralized Finance Liquidations with Reversible Call Options}

\author{Kaihua Qin\inst{1}\inst{4} \and
Jens Ernstberger\inst{2}\inst{4} \and
Liyi Zhou\inst{1}\inst{4} \and
 \\ Philipp Jovanovic\inst{3} \and
Arthur Gervais\inst{3}\inst{4}
}
\authorrunning{K. Qin et al.}
%
\institute{Imperial College London, United Kingdom \and
Technical University Munich, Germany \and
University College London, United Kingdom \and
UC Berkeley RDI, United States}



\maketitle

\begin{abstract}
Liquidations in \DeFi are both a blessing and a curse~---~whereas liquidations prevent lenders from capital loss, they simultaneously lead to \liquidationSpirals and system-wide failures. Since most lending and borrowing protocols assume liquidations are indispensable, there is an increased interest in alternative constructions that prevent immediate systemic-failure under uncertain circumstances.

In this work, we introduce \textit{\financialPrimitives}, a novel financial primitive that enables the seller of a call option to terminate it before maturity. We apply \financialPrimitives to lending in \DeFi and devise \protocol, a protocol for lending platforms to replace the liquidation mechanisms. To the best of our knowledge, \protocol is the first protocol that actively mitigates liquidations to reduce the risk of liquidation spirals. Instead of selling collateral, \protocol incentivizes external entities, so-called \textit{\supporters}, to top-up a borrowing position and grant the borrower additional time to rescue the debt. Our simulation shows that \protocol reduces the amount of liquidated collateral by~\CollateralReduction in a worst-case scenario.




\end{abstract}

\acresetall

\section{Introduction}
Recently, there has been an increasing interest in \DeFi, a financial ecosystem where users exercise cryptographic control over their financial assets.
Commonly, \DeFi is enabled by blockchains that support smart contracts (e.g., Ethereum), and financial primitives are instantiated as publicly accessible decentralized applications. 
A wide variety of traditional financial services that are implemented in \DeFi, ranging from asset exchanges, to market making, as well as lending and borrowing platforms~\cite{qin2021cefi}. 
\DeFi differs from the traditional, centralized financial system in multiple aspects. For instance, most \DeFi services are open-source, such that traders can inspect the protocol rules encoded within immutable smart contracts.

With over~\defiLendingMarketSize of \TVL, \DeFi's lending and borrowing services account for~\defiLendingProportion of \DeFi's locked up assets. Just as in the traditional centralized finance domain, debt in \DeFi is prone to \emph{liquidation events} upon price-swings of the debts' security deposit (subsequently referred to as collateral). A \borrowingPosition becomes ``unhealthy'' (i.e., liquidatable), whenever the collateral is deemed insufficient to cover the debt, corresponding to a \emph{health factor} inferior to one. The most prevalent liquidation mechanism, \FSL, allows a \emph{liquidator} to repay a fraction  of the borrower's debt and acquire its collateral at a discount. The fraction at which the borrowers' debt is repaid in a liquidation is limited to an upper bound, commonly referred to as the \emph{close factor} (e.g.,~\href{https://docs.aave.com/developers/v/2.0/guides/liquidations#0.-prerequisites}{$50\%$}). 
As such, liquidations intend to protect the lender by preventing a loss of capital by selling a sufficient amount of collateral.
However, liquidations serve as a double-edged sword. Selling off collateral causes a price decrease, which potentially leads to further liquidations and market-wide panic~\cite{klages2019stability}.
Quantifying the extent of liquidations in \DeFi, a recent two-year longitudinal study (April $2019$ to April $2021$) by Qin \emph{et al.}~\cite{qin2021empirical} finds that liquidation events on the Ethereum blockchain amount to over $800$M USD in volume, yielding a staggering $64$M USD profit to liquidators. Such liquidation profit constitutes a source of \MEV~\cite{daian2020flash}, which grants miners a risk-free opportunity to extract financial profit. \MEV, however, negatively affects blockchain consensus security by incentivizing blockchain forks~\cite{qin2022quantifying}.


In this work, we propose \protocol, a mechanism designed to mitigate liquidation events to \textit{(i)}~protect borrowers from excessive collateral liquidation, \textit{(ii)}~alleviate \MEV sourcing, and \textit{(iii)}~mitigate \liquidationSpirals. 
To this end, we introduce \textit{\financialPrimitives}, a novel financial primitive that enables the seller of a call option to terminate it at a premium before reaching maturity.
\protocol applies \financialPrimitives to incentivize external support for ``unhealthy'' borrowing positions, while the original borrower is granted additional time to protect its borrowing position and limit the potential loss. 

Thereby, we summarize the contributions of this work as follows.
\begin{enumerate}
    \item \textbf{Quantifying \LiquidationSpiral.} We quantify the \liquidationSpiral caused by the \FSL mechanism by analyzing~\TotalLiquidationEvents past liquidation events over a time-frame of~\liquidationTimeFrame, capturing~\LiquidatedCollateral of collateral liquidated. We find the existence of~\ShortLiquidations short liquidations, where a liquidator immediately sells the acquired collateral.
    These liquidations account for~\ShortLiquidationUSDSold sold collateral and a maximal collateral price decline of~\ShortLiquidationPriceDeclineMax.
    \item \textbf{A Novel Financial Primitive.}
    We introduce \financialPrimitives, a novel financial primitive where the seller of a European call option can pay a premium to the buyer to terminate the option before its maturity.
    \item \textbf{A Protocol for Liquidation Mitigation.} We propose \protocol, the first protocol that protects \DeFi borrowers from excessive liquidation losses. 
    By realizing a \financialPrimitive, \protocol incentivizes external actors to support ``unhealthy'' \borrowingPositions, mitigating liquidations by design. 
    \protocol serves as a plug-and-play mechanism, which can be integrated into any existing lending platform. 
    We evaluate \protocol by simulating how it would have performed in past liquidation events. We find that \protocol reduces the amount of liquidated collateral by~\CollateralReduction in a worst-case scenario.
\end{enumerate}

\section{Background}\label{sec:background}

\subsection{Blockchain \& Smart Contract}
In essence, a blockchain is a distributed ledger operating on top of a peer-to-peer (P2P) network~\cite{bonneau2015sok}. The core blockchain functionality is that participants can transfer financial assets (i.e., cryptocurrencies) without any trusted third-party custodian~\cite{bitcoin}. To send cryptocurrencies, one broadcasts a signed transaction through the blockchain P2P network. The so-called \emph{miners} collect, verify and package transactions into a block which is appended onto the already confirmed blocks forming a linear chain. All peers in the blockchain network are expected to follow a specific consensus mechanism (e.g., Nakamoto consensus~\cite{bitcoin}) to achieve the consistency of the ledger.

Beyond the simple cryptocurrency transfer, more versatile blockchains (e.g., Ethereum~\cite{wood2014ethereum}) enable advanced transaction logic through pseudo-Turing complete smart contracts. Similar to regular user accounts, smart contracts can own cryptocurrencies. In addition, every smart contract is bound to a piece of immutable code upon its creation. Users can send a transaction to a smart contract account and trigger the execution of the associated smart contract code. We refer readers to~\cite{bonneau2015sok} for more detailed explanations of blockchains and smart contracts.

\subsection{Decentralized Finance}
Smart contracts enable the creation of cryptocurrencies (also known as tokens) on a blockchain in addition to the native cryptocurrency (e.g., ETH on Ethereum). A token smart contract serves as a balance sheet recording the balance of every token holder account. Smart contracts also allow anyone to create any type of imaginable financial product on-chain, by enforcing the rules through the smart contracts' immutable code. The ecosystem as a whole, composed of these tokens and smart contract-based financial products, is referred to as \DeFi. At the time of writing, the scale of \DeFi has reached over~\defiMarketSize, with an abundance of applications such as exchanges, lending platforms, and derivatives.\footnote{\url{https://defillama.com/}.}

\subsection{Lending/Borrowing in \DeFi}\label{sec:background-lending-borrowing}
Lending and borrowing, with over~\defiLendingMarketSize \TVL, is one of the most popular \DeFi use cases. In a \DeFi lending system, a smart contract called \emph{lending pool}, manages the \borrowingPositions.
Lenders provide assets to the lending pool to earn interests from borrowers. To minimize the lenders' risk of losing funds, every borrower is required to provide \emph{collateral} as a guarantee.
The lending and borrowing interests are programmatically determined by the contract code.

Lending in \DeFi can be divided into \emph{over-collateralized} and \emph{under-collateralized} lending. In over-collateralized lending, the borrower provides a security deposit (i.e., collateral) which \emph{exceeds} the lent assets by a factor of $1.1\times$ to $2\times$ depending on the respective protocol~\cite{qin2021empirical}. The borrower may then choose to freely use the lent asset in any capacity. Contrary to over-collateralized lending, in under-collateralized lending, the borrower only provides a fraction of the lent assets as security, hence achieving a leverage factor beyond $1\times$. For this leveraged borrowing to remain secure, the assets granted through under-collateralization can only be utilized in very specific, hard-coded settings encoded in immutable smart contracts, such that the lending pools stay in control of the lent assets. In this work, we primarily focus on over-collateralized lending.

We refer to the debts of a borrower together with the collateral securing these debts as a \emph{borrowing position}. Due to asset price fluctuations, the collateral of a borrowing position may become insufficient to cover the debt. Therefore, lending pools typically set a threshold for the borrowing positions, at which a position becomes liquidatable. When the collateral value of a borrowing position declines below this threshold, lending pools can then allow the so-called liquidators, to repay the debt for the position, commonly referred to as liquidation. In return, the liquidator is eligible to acquire parts of the collateral from the borrowing position. The acquired collateral exceeds the repaid debt in value, which incentivizes the liquidator to realize a profit.

\subsection{Call Options}\label{section:callOption}
Call options are financial contracts that grant buyers the right, but not the obligation, to buy an underlying asset (e.g., stocks) at an agreed-upon price (i.e., the exercise price or strike price) and date (i.e., the expiration date or maturity)~\cite{stoll1969relationship,hull2003options}. 
In general, options are priced using a mathematical model, such as the Black-Scholes~\cite{black1973pricing} or the Binomial pricing model~\cite{shreve2005stochastic}.
On a high level, an options price is determined by \emph{(i)} its intrinsic value and \emph{(ii)} its time value.
The intrinsic value is a measure of the profitability of an option if it were to be exercised immediately.
The time value measures the value of an option arising from the time left to maturity (i.e., volatility).
When the strike price of an option increases, the price of the call option consequently increases as well.
In traditional finance, there are two styles of option contracts: \textit{(i)} American options can be executed (or exercised) at any time up to the expiration date; \textit{(ii)} European options can be exercised only on the expiration date~\cite{hull2003options}. 

\section{Preliminaries}\label{sec:preliminaries}

In the following, we formalize a collateralized debt model and the fixed spread liquidation, which is the prevalent \DeFi liquidation mechanism.

\subsection{Collateralized Debt Model}\label{sec:collateralized-debt-model}

We assume the existence of an on-chain lending pool $\lendingPoolShort = \{\borrowingPositionShort{1}, \borrowingPositionShort{2}, ..., \borrowingPositionShort{n}\}$, where $\borrowingPositionShort{i}$ is the $i$-th \borrowingPosition in the lending pool.
Each \borrowingPosition $\borrowingPositionShort{} = \langle \numberCoinsDebt{t}, \numberCoinsCollateral{t} \rangle$ is parametrized by the debt $\numberCoinsDebt{t}$ the borrower owes, and the collateral $\numberCoinsCollateral{t}$ the borrower owns at time $t$.
We denote the price of the debt cryptocurrency towards the collateral cryptocurrency, provided by an oracle~\cite{eskandari2021sok}, as $\price$.
In the following, we consider the case where each \borrowingPosition consists of a single debt cryptocurrency and a single collateral cryptocurrency.
In practice, a \lendingPool may allow for mixed \borrowingPositions by including multiple cryptocurrencies as either debt or collateral.
We further assume that a borrower only opens a single \borrowingPosition.

Whether or not a \borrowingPosition is \textit{liquidatable} is determined by the \textit{\healthFactor}.
\begin{equation}\label{eq:hft}
    \healthFactorShort{\borrowingPositionShort{}}=\frac{\numberCoinsCollateral{t} \cdot \price \cdot \collateralDiscountShort}{\numberCoinsDebt{t}}
\end{equation}
$\numberCoinsCollateral{t} \cdot \price$ represents the value of the collateral, whereas $\numberCoinsDebt{t}$ represents the value of the debt denoted in the same cryptocurrency. $\collateralDiscountShort$ is the collateral discount, s.t. $0 < \collateralDiscountShort < 1$. The \collateralDiscount is configured as a safety margin to ensure the over-collateralization of a position, i.e., the value of the collateral is discounted when calculating the health factor. If $\healthFactorShort{\borrowingPositionShort{}} < 1$, e.g., due to price fluctuations, $\borrowingPositionShort{}$ is deemed ``unhealthy'' making it available for liquidations under existing prevalent designs of \DeFi lending protocols.
Internally, the \healthFactor of a \borrowingPosition relies on the \textit{\collateralizationRatio}

\begin{equation}\label{eq:cr}
    \collateralizationRatioShort{\borrowingPositionShort{}}=\frac{\numberCoinsCollateral{t} \cdot \price}{\numberCoinsDebt{t}}.
\end{equation}
The \collateralizationRatio determines whether a position is over-collateralized or under-collateralized. 
If $\collateralizationRatioShort{\borrowingPositionShort{}} > 1$ at time $t$, a position is over-collateralized, and under-collateralized otherwise.

\subsection{Fixed Spread Liquidation}\label{sec:fixed-spread-liquidation}

We denote a decentralized application for lending and borrowing that applies a fixed spread liquidation mechanism as protocol $\fixedSpreadProtocol$. For ease of exposition, we assume that $\fixedSpreadProtocol$ hosts a single lending pool $\lendingPoolShort$. 
The liquidation of a position $\borrowingPositionShort{} = \langle \numberCoinsDebt{t}, \numberCoinsCollateral{t} \rangle$  is determined by a set of variables, 
including the previously introduced \collateralDiscount $\collateralDiscountShort$, the \closeFactor $\closeFactorShort$ (s.t.\ $0<\closeFactorShort\leq1$) and the  \liquidationSpread $\liquidationSpreadShort$.

\begin{equation}\label{eq:fsl}
    \fixedSpreadProtocol = \langle \lendingPoolShort, \collateralDiscountShort, \closeFactorShort, \liquidationSpreadShort \rangle
\end{equation}

The \closeFactor $\closeFactorShort$ describes the percentage of debt that the liquidator can repay in a single fixed spread liquidation.
The spread $\liquidationSpreadShort$ is the discount at which the liquidator can obtain the collateral. $\liquidationSpreadShort$ is fixed throughout the execution of the protocol (i.e., the name \textit{fixed spread liquidation}).
With the liquidation spread, one can calculate the maximal collateral claimable by the liquidator $\liquidatorShort$ as $(\numberCoinsDebt{t} \cdot \closeFactorShort) \cdot (1+\liquidationSpreadShort)$. 
Without consideration of gas fees, the maximal obtainable profit by $\liquidatorShort$ is $(\numberCoinsDebt{t} \cdot \closeFactorShort) \cdot \liquidationSpreadShort$.
As the protocol is overall a zero-sum game, and under the assumption of non-existant slippage, the profit of the \liquidator is equivalent to the borrowers loss, if denoted in the same cryptocurrency.

Other liquidation mechanisms, though operated differently from the fixed spread liquidations, follow similar high-level designs --- debts are repaid in exchange for collateral from the liquidated borrowing position. For example, in MakerDAO auction liquidations, liquidators bid for the liquidation opportunity by submitting transactions~\cite{qin2021empirical}. In such a setting, the liquidation spread can hence be considered dynamic during the auction execution.

\section{Motivation}\label{sec:motivation}
We proceed to outline the design flaws of liquidation mechanisms and motivate why mitigating liquidations is necessary.

\begin{enumerate}
    \item \textbf{Over-Liquidation.} \DeFi borrowers are exposed to an unnecessarily overwhelming liquidation risk. In regular \FSL configurations,~$50\%\sim 100\%$ of a borrowing position is liquidated within a single transaction~\cite{qin2021empirical,wang2022speculative}.

    \item \textbf{\MEV.} Liquidation is one of the major sources of \MEV, which disrupts miner incentives and endangers the consensus security of a blockchain~\cite{qin2022quantifying}.
    
    \item \textbf{\LiquidationSpiral.} A liquidation increases the supply of the collateral cryptocurrencies available for sale. This supply inflation imposes a negative impact on the collateral prices~\cite{whelan2001economic} and may result in further liquidations (possibly \liquidationSpiral~\cite{klages2019stability}). We provide a case study of a real liquidation event to present the impact of liquidations on collateral prices.
\end{enumerate}



\begin{figure}[h]
    \centering
    \includegraphics[width=\columnwidth]{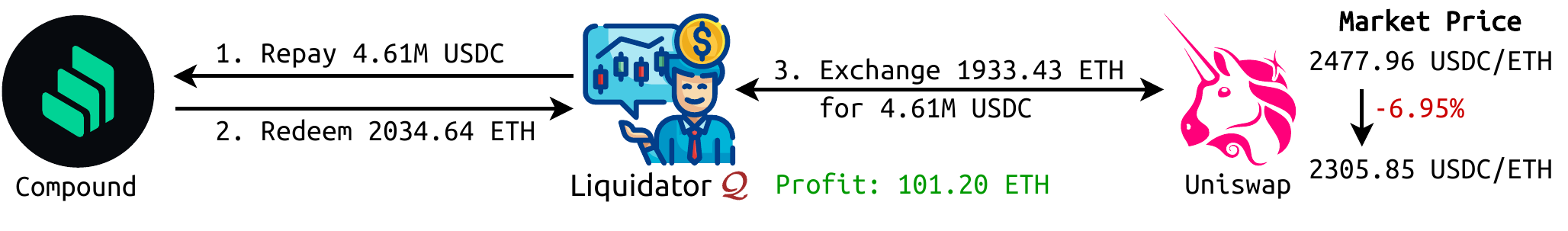}
    \caption{A real liquidation event with a subsequent downward price trend of the collateral asset. The liquidator $\liquidatorShort$ immediately sold parts of the redeemed ETH collateral from a Compound liquidation, which decreased the ETH price on Uniswap by $6.95\%$.}
    \label{fig:liquidation-case-study}
\end{figure}

\newtheorem{casestudy}{Case Study}
\begin{casestudy}[\LiquidationSpiral]\label{casestudy:deleveraging-spiral}
As shown in Figure~\ref{fig:liquidation-case-study}, two \DeFi platforms were involved in this liquidation: (i) Compound, an over-collateralized lending platform; (ii) Uniswap, an on-chain exchange. USDC is a stablecoin, of which the value is pegged to USD.\footnote{Transaction hash: \etherscantx{0xe7b6fac6502be7c6659880ff5d342ec470429c6f49cd457945bf0726667eb689}. Note that we ignore the irrelevant execution details to ease understanding.}  In the studied liquidation, the liquidator mainly took the following three steps.

\begin{enumerate}
    \item The liquidator repaid~$4.61$M USDC for a Compound borrowing position.
    \item In return, the liquidator was allowed to redeem $\numprint{2034.64}$ ETH of collateral.
    \item The liquidator bought $\numprint{1933.43}$ ETH from the redeemed collateral and exchanged the ETH for $4.61$M~USDC to cover its repayment cost in Step 1. The liquidator realized a profit of~$101.20$ ETH through this liquidation.
\end{enumerate}
In the third step, the exchange from ETH to USDC on Uniswap USDC/ETH triggered a price decline from~$\numprint{2477.96}$ USDC/ETH to~$\numprint{2305.85}$ USDC/ETH ($-6.95\%$). This event shows that even a single liquidation can decrease the collateral price significantly.
\end{casestudy}

\subsubsection{Why collateralization instead of liquidation?}
In this work, \protocol requires additional collateral to be locked in the lending pool, reducing the liquid collateral asset supply. Hence, we conclude that \protocol behaves more positively than a liquidation mechanism on stabilizing lending markets, effectively acting like a price ``softening buffer''. We empirically confirm this effect in Section~\ref{sec:empirical}.

\section{\protocol}\label{sec:toll}

In this section, we introduce \protocol. \protocol is a debt management mechanism for \DeFi lending protocols. 
It mitigates liquidations through a set of incentives that decrease the likelihood of \liquidationSpirals.  
\protocol relies on \textit{\supporters}, which are entities that are willing to top up unhealthy borrowing positions. To enable \protocol, we introduce \textit{\financialPrimitives}, a novel financial primitive where the seller of a call option can pay a premium to terminate the contract before maturity.

\subsection{\FinancialPrimitive}

Recall the notion of European call options as introduced in Section~\ref{section:callOption}.
In a European call option, the seller offers the option contract whereas the buyer acquires the option to exercise the right to buy an asset at a specific price by buying said option at a premium (i.e., the option price).
The outcome of a European call option contract at maturity is binary~---~\emph{(i)} the buyer exercises its right to buy or \emph{(ii)} the buyer does not exercise its right to buy.
We now introduce the \textit{reversible} European call option, which augments the traditional European call option with an additional outcome to the option contract, where the seller is able to terminate the contract at a premium. 

We say that a reversible European call option contract gives the buyer $\buyerShort$ the option, but not the obligation, to buy a specified amount $\amountAssetShort$ of an asset $\AssetShort$ at a specified price $\strikePriceShort$ at maturity $\terminationTime$, and the seller $\sellerShort$ the option to terminate the contract at pre-Maturity $\initiateTime < \Time < \terminationTime$.
The buyer $\buyerShort$ pays a premium $\premiumShort$ at $\initiateTime$ for the option to exercise the contract at maturity $\terminationTime$.

Formally, we define \financialPrimitive as follows:

\begin{definition}[\FinancialPrimitive]
    A \financialPrimitive is parameterized by an asset $\AssetShort$, the asset amount $\amountAssetShort$, the strike price $\strikePriceShort$, the reimbursement factor $\factorPrimitive$, and the time of maturity $\terminationTime$.
    The mechanics are as follows:
    
    $\initiateTime$: 
    \emph{(i)} The contract is agreed upon between $\buyerShort$ and $\sellerShort$. \newline
    \mainIndent \emph{(ii)} The \buyer $\buyerShort$ pays a premium $\premiumShort$ to the \seller $\sellerShort$. 

    $\initiateTime < \Time < \terminationTime$:
    The \seller $\sellerShort$ of the option can choose to terminate the contract \newline
    \mainIndentTwo by reimbursing the buyer $\buyerShort$ with $\premiumShort \cdot \factorPrimitive$. 

    $\terminationTime$:
    The \buyer $\buyerShort$ can acquire $\amountAssetShort$ units of asset $\AssetShort$ at  strike price~$\strikePriceShort$.
    
\end{definition}





\begin{figure}[!t]
    \centering
    \begin{tikzpicture}[
      declare function={
        mypl(\x)= -1.25 + (\x>25) * (0.9*(\x-25));
      }
    ]
    \begin{axis}[
      xlabel={Price $A$},
      ylabel={$P\&L$},
      axis lines=middle,
      width=10cm,height=5cm,
      xmin=0,xmax=4.9,
      ymin=-2,ymax=2.5,
      ytick={-1,1.5}, 
      xtick={2, 3},
      xticklabels={$\strikePriceShort$, $\strikePriceShort+\premiumShort$},
      yticklabels={$-\premiumShort$, $\premiumShort \cdot \factorPrimitive$},
      legend style={
        draw=none,
        legend columns=-1,
        at={(0.5,1)},
        anchor=south,
        outer sep=1em,
        node font=\small,
      },
    ]
    
      \addplot[blue,thick] coordinates {(0,1.5) (5,1.5)};
      \addlegendentry{P\&L at reversion}
      \addplot[red,thick] coordinates {(0,-1) (2,-1)};
      \addplot[red,thick] coordinates {(2,-1) (5,2)};
      \addlegendentry{P\&L at Maturity};
    \end{axis}
    \end{tikzpicture}
    \caption{Payoff \& Loss (P\&L) analysis for the Buyer $\buyerShort$ of the \financialPrimitive. In case of reversion, the payoff for the $\buyerShort$ is constant. In case of maturity, the payoff is equal to a traditional call option.}
    \label{figure:payoffCurve}
\end{figure}

\noindent \textbf{Payoff analysis.}
The buyer $\buyerShort$ is the entity which is entitled to execute the option contract at maturity.
We assume that $\buyerShort$ always acts rationally, such that their financial benefit is maximized. 
In the case of a \financialPrimitive, the payoff which $\buyerShort$ receives can be categorized into two cases~---~\emph{(i)} $\sellerShort$ terminates the option at pre-maturity or \emph{(ii)} the contract is not terminated until maturity at time $\terminationTime$.
In the first case, the payoff for $\buyerShort$ is constant, as the seller $\sellerShort$ reimburses the buyer $\buyerShort$ with $\premiumShort \cdot \factorPrimitive$, where $\factorPrimitive > 1$. 
If the seller $\sellerShort$ does not terminate the contract, the payoff for $\buyerShort$ is equivalent to
\begin{equation}
    P_{\buyerShort} = 
    \begin{cases}
      A(T)-K-\premiumShort & \text{if~} A(T) \geq K \\
      -\premiumShort & \text{if~} A(T) < K
    \end{cases}
\end{equation}
Note, that the payoff in this case is equivalent to a traditional European style call option.
The visualized payoff curves for $\buyerShort$ are presented in Figure~\ref{figure:payoffCurve}.


\subsection{The \protocol Protocol}

We present the \protocol protocol in the following. On a high-level, \protocol seeks to mitigate liquidations through \supporters that top-up the collateral of an unhealthy borrowing position (i.e., the health factor declined below one). 
\protocol allows any external entity to become such a \supporter.
We start with an overview of \protocol by outlining the equivalence to \financialPrimitives.

\subsubsection{Overview}

\begin{figure}[tb!]
    \centering
    \includegraphics[width=\columnwidth]{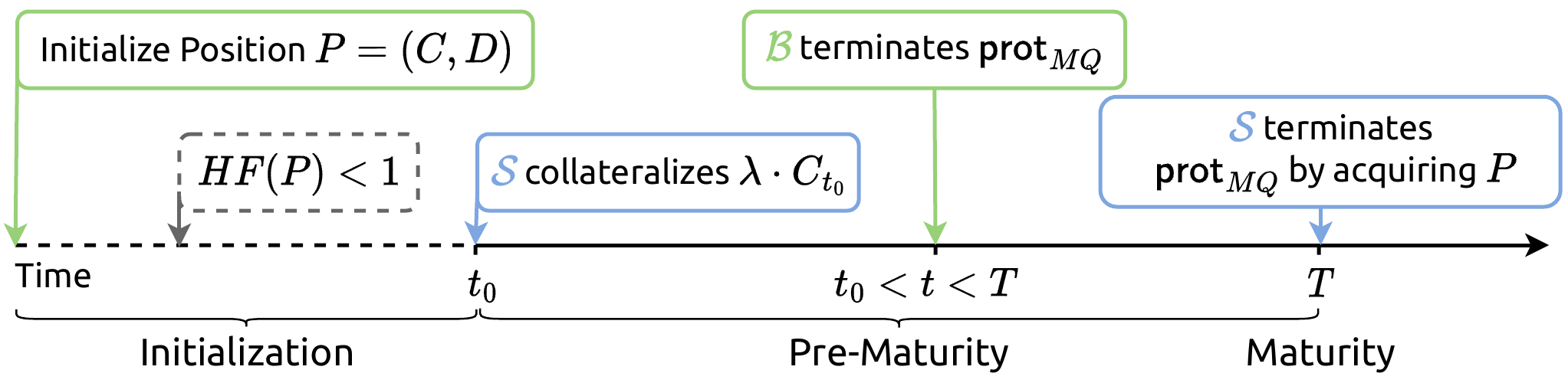}
    \caption{High-level overview of the \protocol protocol which realizes a reversible call option in \DeFi.
    Once the borrowing position opened by the \borrower $\borrowerShort$ is unhealthy, yet not liquidated, the \supporter $\supporterShort$ is able to top up the collateral in position $\borrowingPositionShort{}$.}
    \label{fig:high-level-toll}
\end{figure}


An overview of \protocol is presented in Figure~\ref{fig:high-level-toll}.
On a high-level, \protocol is separated into three phases~---~\emph{(i)} Initialization, \emph{(ii)} pre-Maturity, and \emph{(iii)} Maturity.
We first assume that \protocol replaces the liquidation mechanism in our exemplary lending/borrowing protocol. We defer practical considerations for co-existence of \protocol and liquidations to Section~\ref{subsection:practicalInstantiation}.

\begin{description}[style=unboxed, leftmargin=0cm]
    \item[1) Initialization.]
    We assume the existence of an on-chain \lendingPool $\lendingPoolShort$ with a single borrowing position $\borrowingPositionShort{} = \langle \numberCoinsDebt{t}, \numberCoinsCollateral{t} \rangle$ initialized by the \borrower $\borrowerShort$.
    The supporter can engage at time $\initiateTime$, if the following condition holds:
    
    \begin{equation}
    \label{equation:HF}
    \healthFactorShortTwo{\initiateTime}{\borrowingPositionShort{}} = 
    \frac{\numberCoinsCollateral{\initiateTime} \cdot \priceInp{\initiateTime} \cdot \collateralDiscountShort}{\numberCoinsDebt{\initiateTime}} < 1
    \end{equation}


In words, the \healthFactor should be lower than one. Note that the position may be over-collateralized ($\collateralizationRatioShortTwo{\initiateTime}{\borrowingPositionShort{}} > 1$) or under-collateralized  ($\collateralizationRatioShortTwo{\initiateTime}{\borrowingPositionShort{}} < 1$), depending on the steepness of the price decline that yields a \borrowingPosition unhealthy.
At this point, $\supporterShort$ buys a \financialPrimitive by topping-up $\factorMiqado \cdot \numberCoinsCollateral{\initiateTime}$ into $\borrowingPositionShort{}$, which grants the right to take over the \borrowingPosition $\borrowingPositionShort{}$ at maturity $\terminationTime$.
The price of the \financialPrimitive hence is $\factorMiqado \cdot \numberCoinsCollateral{\initiateTime}$. Note that the premium factor $\factorMiqado$ is a protocol parameter that can be ruled in the lending pool contract. To decide whether to deposit, a supporter would need to price the \financialPrimitive and estimate its potential profitability, which we detail in Section~\ref{sec:pricing}.

\item[2) pre-Maturity.]
Once $\supporterShort$ acquires a \protocol option with maturity $\terminationTime$, the pre-maturity stage starts.
At any point $\initiateTime < \Time < \terminationTime$, the borrower $\borrowerShort$ can terminate the \protocol protocol by repaying $\supporterShort$ the premium $\factorMiqado \cdot \numberCoinsCollateral{\initiateTime}$ multiplied by a constant factor $\factorMiqadoRe$ that incentivizes the initial support of $\supporterShort$, hence 

\begin{equation}
    \returnPaymentShort = \factorMiqado \cdot \numberCoinsCollateral{\initiateTime} \cdot (1+\borrowingInterestRate) \cdot \factorMiqadoRe 
\end{equation}

where $0 < \borrowingInterestRate < 1$ is the interest rate which $\borrowerShort$ agreed to pay for its loan when initiating the position $\borrowingPositionShort{}$.
The factor $0 < \factorMiqadoRe < 1$ is implementation dependent and should account for the risk $\supporterShort$ has to take when supporting a position.
    
\item[3) Maturity.]
At Maturity, there are two possible options how the \protocol protocol may terminate. The payoff for the \supporter $\supporterShort$ in the case of maturity is depicted in Figure~\ref{figure:payoffCurve}.
\begin{enumerate}
    \item \textbf{Full Takeover.}
    In general, \protocol option contracts have an ``Out-of-the-Money'' strike price $\strikePriceShort$, such that the strike is greater than the collateralization ratio upon initiation of the position $\borrowingPositionShort{}$ by $\borrowerShort$. Essentially, as the health factor is lower than one, the intrinsic value of the option is low, whereas the time value based on volatility and time of expiration is high.

    \item \textbf{Default.}
    The \supporter defaults and does not exercise the option, hence loses the premium $\premiumShort$ (cf. Figure~\ref{figure:payoffCurve}), if the price at Maturity is below the strike price $\strikePriceShort$. In this case, where \protocol fully replaces the liquidation mechanism, another round of Miqado initiates. Rational \supporters initiate a \protocol session if the condition presented in \emph{1.) Initialization} is fulfilled.
    
\end{enumerate}
    
\end{description}


\subsubsection{Incentive Discussion}

A common question is why a \supporter would actually engage in the \protocol protocol and top up liquidity positions that are unhealthy.
In general, whether a \supporter is incentivized to engage in a \protocol option in a \FSL liquidity pool depends on the price volatility and the selected strike price.
Given the volatility of various cryptocurrencies, it is infeasible to draw a general conclusion fitting all scenarios.
\Supporters can price the \protocol options and compare to the required cost (i.e., the premium) to evaluate the potential risks. We outline a pricing model for \financialPrimitives in Section~\ref{sec:pricing}.
In practice, we assume that \supporters taking a low risk will face termination at pre-maturity by $\borrowerShort$, yielding a smaller payoff for $\supporterShort$.
We empirically evaluate \protocol's ability to prevent \liquidationSpirals by replacing the liquidation mechanism in Section~\ref{sec:empirical}.





%



\subsection{Pricing \FinancialPrimitive}\label{sec:pricing}
The \financialPrimitive is equivalent to an European call option in the case of maturity. Therefore, we can apply the widely adopted Black-Scholes pricing model~\cite{hull2003options} for European call options to \protocol.
We outline the B-S model details in Appendix~\ref{app:bs-model}.
We assume that at initialization $\initiateTime$, the \supporter $\supporterShort$ buys a \protocol option by supplying $\factorMiqado \cdot \numberCoinsCollateral{\initiateTime}$ of additional collateral priced at $\factorMiqado \cdot \numberCoinsCollateral{\initiateTime} \cdot \priceInp{\initiateTime}$.
    The spot exchange rate is equivalent to $\priceInp{\initiateTime}$, whereas the domestic interest rate $\domesticInterestRate$ is equivalent to the borrowing interest rate of the protocol $\borrowingInterestRate$.
    The foreign interest rate $\foreignInterestRate$ remains the same.
    The volatility $\volatilityAsset$ can be calculated from the price history.
    Henceforth, the optimal factor $\factorMiqado^{*}$ following the B-S model can be calculated as
    \begin{equation}
    \label{equation:price}
        \factorMiqado^{*} = \frac{\priceInp{\initiateTime} e^{-\foreignInterestRate \cdot \terminationTime} N(d_1) - \strikePriceShort e^{-\borrowingInterestRate \cdot \terminationTime} N(d_2)}{\numberCoinsCollateral{\initiateTime} \cdot \priceInp{\initiateTime}}
    \end{equation}
    with equations for $d_1$ and $d_2$ outlined in Appendix~\ref{app:bs-model}. A \supporter then compares the actual premium factor $\factorMiqado$ set by the lending protocol to $\factorMiqado^{*}$ and evaluates the profitability. In practice, a supporter would have a personalized pricing model based on the supporter's predictions on the price dynamics and risk preference.
    

\subsection{Practical Instantiation}
\label{subsection:practicalInstantiation}

\begin{figure}[tb!]
    \centering
    \includegraphics[width=0.75\columnwidth]{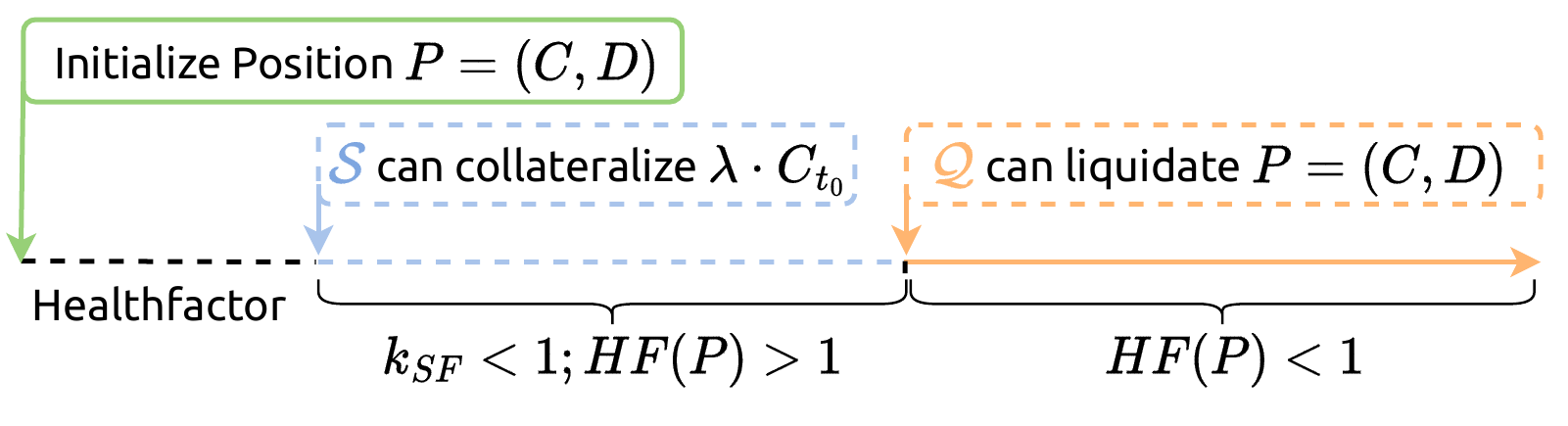}
    \caption{
    Practical Instantiation of \protocol on top of a traditional liquidation mechanism.
    The \supporter $\supporterShort$ has an advantage over the liquidator $\liquidatorShort$ to support a temporarily unhealthy position.
    }
    \label{fig:practical-instantiation}
\end{figure}

When there is no \supporter $\supporterShort$ willing to purchase a \financialPrimitive or when a \supporter defaults, the \lender $\lenderShort$ faces a loss as the \borrower  $\borrowerShort$ is not incentivized to repay the outstanding debt and $\supporterShort$ is not incentivized to take over the position $\borrowingPositionShort{}$.
In a practical instantiation (cf.\ Figure~\ref{fig:practical-instantiation}), a protocol operator may want to operate \protocol options on top of a traditional liquidation mechanism in order to prevent this.
As such, the protocol can employ a buffer to derive an additional \textit{support factor} $\supportFactorShort$, such that $\supporterShort$ can engage in a \protocol option if 
\begin{equation}
    \supportFactorShort = \collateralizationRatioShortTwo{\initiateTime}{\borrowingPositionShort{}} \cdot (\collateralDiscountShort + \bufferShort) < 1
\end{equation}
where $\bufferShort$ is the buffer parameter, s.t.\ $\bufferShort > 1$.

A liquidator can additionally engage when the health factor is lower than one, as traditionally assumed and presented in Equation~\ref{equation:HF}.
With this construction, the \supporter has an advantage over the \liquidator to support a temporarily unhealthy position and make a profit.
Effectively, this construction similarly mitigates \liquidationSpirals, dependent on the buffer $\bufferShort$.

\subsection{Remarks}
\protocol enhances Fixed Spread Liquidations in the following aspects:
\begin{description}[style=unboxed, leftmargin=0cm]
    \item[Rescue Opportunity.] The reversible call option of \protocol offers a time window for a borrower to rescue its borrowing position. With a fixed spread liquidation, the close factor is usually larger than necessary such that more collateral is sold off at a discount, which negatively impacts the borrowers financial interests. With \protocol options, this risk is alleviated, such that over-liquidation is not a concern and the \borrower has to pay less to rescue its position.
    \item[Collateral Restraint.] \protocol absorbs additional collateral and locks it in the lending pool until the reversible call option's maturity. This mitigates the possible \liquidationSpiral, which we quantitatively show in Section~\ref{sec:empirical}.
    \item[\MEV Mitigation.] \FSL liquidations provides deterministic and cost-free  opportunities for miners to profit through manipulating transaction order and front-running other liquidators. In \protocol, if a miner deems a reversible call option profitable, it still has an advantage over other supporters. This is because a miner can single-handedly front-run any competing transaction and be the first to initiate \protocol. Nevertheless, as shown in Section~\ref{sec:protocol-evaluation}, a \protocol reversible call option does not guarantee a profit. Moreover, a supporter bears a capital cost while locking the premium in the lending pool. We hence conclude that \protocol mitigates the \MEV problem.
\end{description}

\section{Empirical Evaluation}\label{sec:empirical}
In this section, we evaluate the \protocol protocol by comparing \protocol to the dominant liquidation mechanism \FSL. To this end, we collect all liquidation events on Aave (both V$1$ and V$2$) and Compound from the~\StartDate to the~\EndDate. Aave and Compound are the top two lending protocols on Ethereum in terms of \TVL, according to \href{https://defillama.com/protocols/lending/Ethereum}{defillama.com}. Both of the two lending protocols follow the \FSL mechanism (cf.\ Section~\ref{sec:fixed-spread-liquidation}). In total, we collect~\TotalLiquidationEvents liquidations (Aave V1:~\AaveVOneLiquidations; Aave V2:~\AaveVTwoLiquidations; Compound:~\CompoundLiquidations). 

\subsection{Quantifying \LiquidationSpiral}\label{sec:quantifying}
\subsubsection{Collateral Release.}

A lending protocol that applies \FSL directly sells the liquidated collateral to the \liquidator at a discount. 
This aggravates the price downtrend of the liquidated cryptocurrency as \liquidators may immediately sell of the acquired collateral, which was locked in the lending protocol, to secondary markets.
Precisely measuring the impact of \FSL on the liquidated collateral price is challenging. We need to devise an accurate economic model to exclude the impact of other factors, such as the demand change for the collateral. We also need to model the liquidity dynamics on various centralized and decentralized exchanges at the time of liquidation. These challenges are however beyond the scope of this study and are left for future work. Therefore, we choose to present the value of collateral that is released in the \FSL liquidations (cf.\ Metric~\ref{metric:release}) to intuitively quantify the \liquidationSpiral introduced by the \FSL mechanism.

\begin{figure}[t]
    \centering
    \includegraphics[width=\columnwidth]{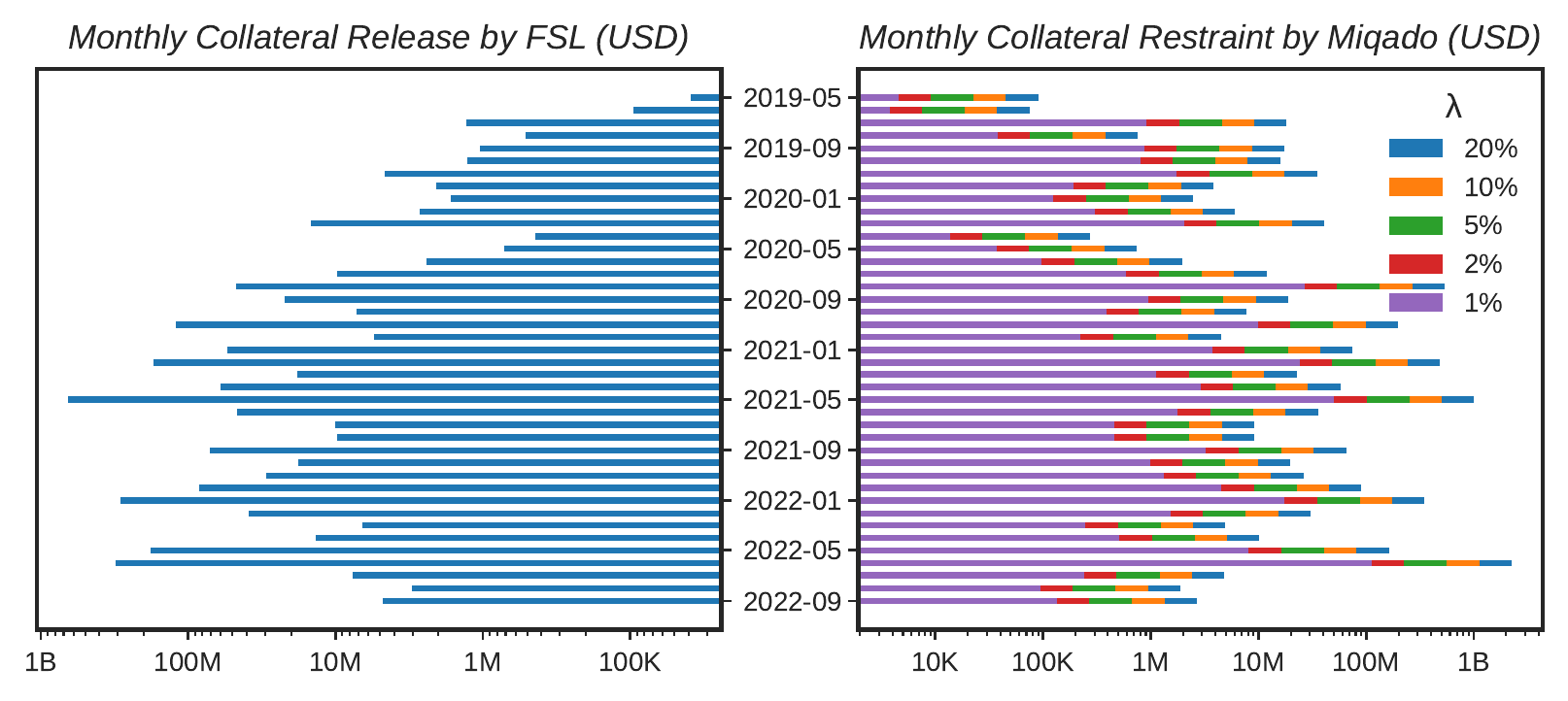}
    \caption{Over a time-frame of~\liquidationTimeFrame (from the~\StartDate to the~\EndDate), the collateral release by the \FSL mechanism accumulates to~\LiquidatedCollateral, with a monthly peak of~\CollateralReleasePeak in May,~$2021$. On the contrary, our \protocol protocol restrains additional collateral in the lending pool instead of releasing and further mitigates the \liquidationSpiral. The accumulative collateral restraint by \protocol (cf.\ Metric~\ref{metric:restraint}, Section~\ref{sec:protocol-evaluation}) amounts to~\CollateralRestraintTwenty~USD when the premium factor $\lambda$ is set to~$20\%$.}
    \label{fig:collateral_release}
\end{figure}

\newtheorem{metric}{Metric}
\begin{metric}[\FSL Collateral Release]\label{metric:release}
The value of collateral released to the markets in a \FSL liquidation.
\end{metric} 
Figure~\ref{fig:collateral_release} presents the monthly collateral release in the past~\TotalLiquidationEvents \FSL liquidations. The total collateral release amounts to~\LiquidatedCollateral over the~\liquidationTimeFrame.

\subsubsection{Direct Price Decline.}
In Case Study~\ref{casestudy:deleveraging-spiral} (cf.\ Section~\ref{sec:motivation}), we show that a liquidator can choose to sell the collateral acquired from the borrower within the liquidation transaction. We observe that such a ``sell-after-liquidation'' strategy is prevalent, which we define as a short liquidation (cf.\ Definition~\ref{def:short-liquidation}). 
\begin{definition}[Short Liquidation]\label{def:short-liquidation}
In a short liquidation, $\liquidatorShort$ sells (fully or partially) the collateral acquired from $\borrowerShort$ within the liquidation transaction.
\end{definition}
To identify a short liquidation, we first gather the ERC-20 transfer and asset swap events from a liquidation transaction.\footnote{ERC-20 is a fungible token standard, which is extensively adopted in the Ethereum \DeFi ecosystem. An event refers to a log emitted by a smart contract during its execution. These events are identifiable by a unique topic hash and can represent various actions, such as an asset swap on a decentralized exchange. In this work, for asset swap events, we captured the most liquid exchanges on Ethereum including Uniswap V1, V2, V3, Sushiswap, and Curve.} With these events, we then filter the exchange contracts that are potentially used for collateral selling. The filtering process is based on two criteria: \textit{(i)} the contract emits an asset swap event during the transaction execution; \textit{(ii)} the contract receives the liquidated collateral token (fully or partially). If such an exchange contract is detected, the liquidation transaction is classified as a short liquidation.
From the~\TotalLiquidationEvents studied liquidations, we identify~\ShortLiquidations short liquidations. In total,~\ShortLiquidationUSDSold of collateral is sold directly by the liquidators in these short liquidations. We find that in~\FullySoldShortLiquidations of the short liquidations, the acquired collateral is fully sold. On average,~\ShortLiquidationsSellPercentageAverage of the collateral is sold in a short liquidation.

A short liquidation directly leads to a collateral price decline on the exchange where the liquidator sells the acquired collateral. Although a significant price change in a single market will eventually be evened out by arbitrageurs\footnote{Entities who profit by leveraging price differences across different markets.} among all available markets, while the negative impact on the collateral price remains. We therefore apply such a price decline as a metric of how \FSL liquidations destabilize lending protocols (cf.\ Metric~\ref{metric:price-decline}).

\begin{metric}[Direct Price Decline]\label{metric:price-decline}
In a short liquidation, the spot price decline on the exchange where the liquidator sells the acquired collateral.
\end{metric}
We find that the average collateral price decline led by the~\ShortLiquidations short liquidations is~\ShortLiquidationPriceDeclineAverage, while the maximal decline reaches~\ShortLiquidationPriceDeclineMax.\footnote{Cf.\ \etherscantx{0xff2d484638b846a46b203a22b02d71df44bf78346c72b954ad0ad05f34b134c8}}

\subsection{\protocol Evaluation}\label{sec:protocol-evaluation}
In the following, we assume that Aave and Compound had adopted \protocol and simulate how \protocol could have outpaced \FSL in the past liquidation events. Our simulation is constrained to every single liquidation event, while ignoring the long-term impact of \protocol. For example, \protocol mitigates the price downtrend and hence could have prevented follow-up liquidations in a \liquidationSpiral, which we leave for future research.

The performance of \protocol is influenced by its parameters. In our simulation, we assume that \protocol follows the corresponding lending protocol's configuration for the collateral discount $\collateralDiscountShort$ at the time of each liquidation. This implies that \protocol shares the same triggering condition as \FSL (i.e., when the health factor declines below one) and hence applies to every liquidated borrowing position. We also need to parameterize the premium factor $\lambda$ and the time to maturity $\Delta T$ for the reversible call option. Similar to how the parameters for lending protocols evolve,\footnote{\url{https://docs.aave.com/risk/asset-risk/risk-parameters}.} these two parameters need to be empirically determined and dynamically adjusted given various market conditions (e.g., the price volatility). We therefore simulate on various specific settings to show how \protocol performs under different configurations.

\subsubsection{Collateral Restraint.} \protocol absorbs additional collateral, which is restrained in the lending pool during the protocol execution. This collateral restraint, contrary to \FSL's supply release (cf.\ Metric~\ref{metric:release}), imposes a positive impact on stabilizing collateral price (cf.\ Metric~\ref{metric:restraint}).
\begin{metric}[\protocol Collateral Restraint]\label{metric:restraint}
The value of collateral deposited by the \supporter in a \protocol execution.
\end{metric}
We visualize the monthly comparison between the collateral restraint by \protocol and the collateral release by \FSL in Figure~\ref{fig:collateral_release}. The accumulative collateral restraint with different parameters is outlined in Table~\ref{tab:accumulative_collateral_restraint}, Appendix~\ref{app:tables}. We find that when $\lambda$ is~$20\%$, the accumulative collateral restraint reaches~\CollateralRestraintTwenty~USD. Notably, as a by-product, the restrained additional collateral is counted towards the lending pool's \TVL, which is a common protocol success metric.

\subsubsection{Health Factor Recovery.} 
\begin{figure}[t]
    \centering
    \includegraphics[width=\columnwidth]{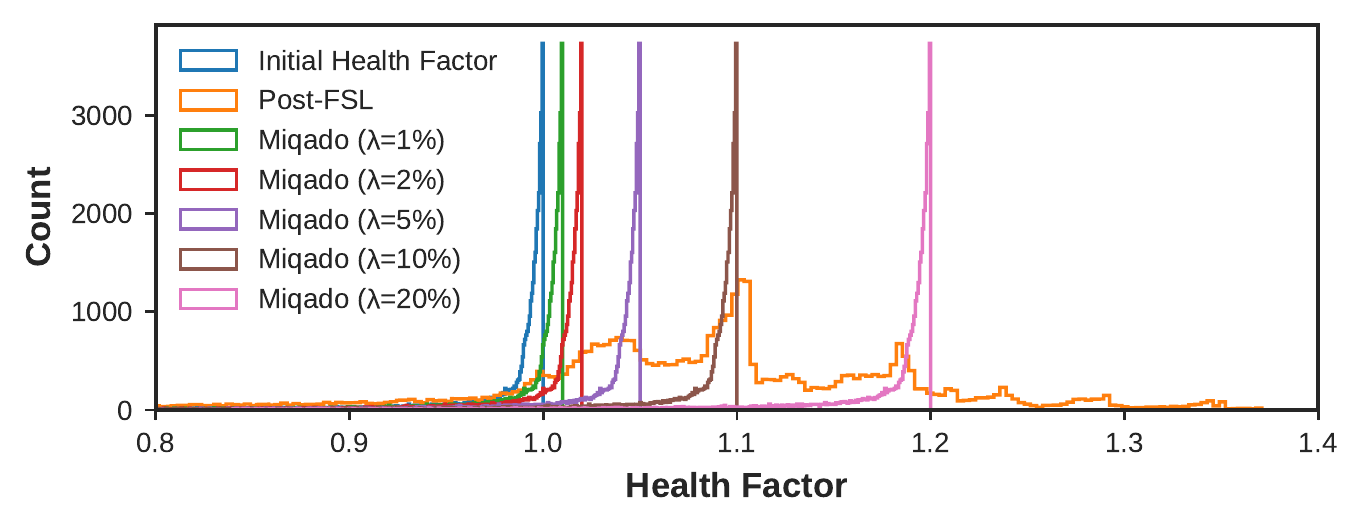}
    \caption{The health factor distributions pre- and post-\FSL liquidations. We also visualize how \protocol increases the health factor with different premium factors.}
    \label{fig:health_factor_distributions}
\end{figure}

One shared target of \protocol and \FSL is to increase the health factor of a borrowing position. In Figure~\ref{fig:health_factor_distributions}, we present the health factor distributions before and after the studied \FSL liquidations. We further simulate how \protocol could have increased the health factor with different parameters. We find that,~\HealthyPositionsAfterFSL of the liquidated positions become healthy (the health factor is increased above one) after a \FSL liquidation. When $\lambda$ is set to $5\%$, \protocol achieves the same performance (\HealthyPositionsAfterMiqadoLambdaFive~of the borrowing positions become healthy after the supporter deposits). 

\subsubsection{Payoffs for \Supporter.} We proceed to simulate the payoffs of \protocol supporters. 
In this section, we assume that the borrowers would not terminate the reversible call options. We parameterize $\Delta T$ to $1$, $6$, and $24$ hours and apply the real market price to value every reversible call options at maturity. A supporter then chooses to exercise the option when the value of collateral exceeds the outstanding debt at maturity, and defaults otherwise (cf.\ Figure~\ref{figure:payoffCurve}). In Table~\ref{tab:payoffs}, Appendix~\ref{app:tables}, we outline the probability that a supporter \textit{(i)} exercises the call option and profits, \textit{(ii)}  exercises the call option but loses, \textit{(iii)} defaults, under different parameters. We also present the average profit for every supporter. We show that, to our surprise, the \protocol premium factor does not impact the probability of the reversible call option in practice. Notably, in Table~\ref{tab:payoffs}, we assume that the borrowers would not rescue their debts and therefore conjecture that the actual payoffs for supporters would be lower than the presented results.

\subsubsection{Collateral Release Reduction} In practice, the probability that a \protocol supporter may default on the reversible call option is up to~$13.48\%$. This implies that the associated borrowing position is under-collateralized at maturity and may be further available for \FSL (cf.\ Section~\ref{subsection:practicalInstantiation}). We simulate that, in the worst case, the collateral release by \FSL after \protocol (cf.\ Metric~\ref{metric:release}) amounts to~\CollateralReleaseWithMiqado, which is a reduction of~\CollateralReduction compared the~\LiquidatedCollateral collateral release by \FSL only (cf.\ Section~\ref{sec:quantifying}).

\section{Related Work}\label{sec:related-work}
Various works in \DeFi focus on lending \& borrowing protocols from diverse perspectives such as economics, security and formal modeling.
Kao \etal~\cite{kao2020analysis} evaluate the economic security of Compound by using agent-based simulation. Darlin \etal~\cite{darlin2020optimal} investigate the optimal bidding strategies for auction liquidations. Perez \etal~\cite{perez2021liquidations} present an empirical analysis of liquidations on Compound. Qin \etal~\cite{qin2021empirical} perform a longitudinal study on the liquidation events of four major Ethereum lending pools (i.e., Aave, Compound, dYdX, and MakerDAO), while showing the over-liquidation problem of the fixed spread liquidations. In this work, we show that the proposed \protocol protocol mitigates these problems. Bartoletti~\etal systematize \DeFi lending pools~\cite{bartoletti2021sok} and further provide a formal analysis of \DeFi lending pools~\cite{bartoletti2022formal}. Wang \etal~\cite{wang2022speculative} study under-collateralized \DeFi lending platforms showing the three main risks of a leverage-engaging borrower, namely, impermanent loss, arbitrage loss, and collateral liquidation. Select stablecoin designs leverage lending and borrowing mechanisms (e.g., DAI from MakerDAO), as studied in~\cite{klages2022while,klages2020stablecoins,klages2019stability}.

Besides \DeFi lending and borrowing, further studies focus on decentralized exchanges and the security of the \DeFi ecosystem~\cite{daian2020flash,zhou2021high,qin2021attacking,qin2022quantifying,zhou2022sok}. 
Most recently, Zhou~\etal~\cite{zhou2022sok} systematize attacks on \DeFi and highlight the need for further research on the protocol layer due to $59$\% of attacks on lending \& borrowing platforms yielding from insufficient protocol design. 

Further, there are various non-academic works that offer call options in decentralized applications.
\href{https://www.hegic.co/}{Hegic} offers gas-free option trading for ETH and BTC. \href{https://www.ribbon.finance/}{Ribbon} supports on-chain options, where the option price, or premium, is set through an auction. However, none of the existing decentralized applications applies an equivalent financial primitive to lending \& borrowing platforms to mitigate liquidations.

\section{Conclusion}\label{sec:conclusion}

We presented \protocol, the first liquidation mitigation protocol. 
Whereas existing lending and borrowing protocols rely on plain liquidation mechanisms, \protocol secures \borrowingPositions by incentivizing external entities to provide additional collateral.
To facilitate \protocol, we introduce \financialPrimitives, a novel financial primitive with promising properties for application in \protocol.
To highlight the need for \protocol, we show that fixed spread liquidations trigger \liquidationSpirals and destabilize lending markets.
We evaluate \protocol by executing \protocol logic on past blockchain states. We show that by applying \protocol, the amount of liquidated collateral can be reduced by~\CollateralReduction.
By providing a plug-in replacement to existing liquidation mechanisms, \protocol can prevent systemic-failures without extensive overhead.

\subsubsection*{Acknowledgements} We thank the anonymous reviewers for the thorough reviews and helpful suggestions that significantly strengthened the paper. This work is partially supported by Lucerne University of Applied Sciences and Arts, the Federal Ministry of Education and Research of Germany (in the programme of ``Souverän. Digital. Vernetzt.''. Joint project 6G-life, project identification number: 16KISK002), and the Algorand Centres of Excellence programme managed by Algorand Foundation.

\bibliographystyle{splncs04}
\bibliography{references.bib}

\begin{thebibliography}{10}
\providecommand{\url}[1]{\texttt{#1}}
\providecommand{\urlprefix}{URL }
\providecommand{\doi}[1]{https://doi.org/#1}

\bibitem{bartoletti2022formal}
Bartoletti, M., Chiang, J., Junttila, T., Lluch~Lafuente, A., Mirelli, M.,
  Vandin, A.: Formal analysis of lending pools in decentralized finance. In:
  Leveraging Applications of Formal Methods, Verification and Validation.
  Adaptation and Learning: 11th International Symposium, ISoLA 2022, Rhodes,
  Greece, October 22--30, 2022, Proceedings, Part III. pp. 335--355. Springer
  (2022)

\bibitem{bartoletti2021sok}
Bartoletti, M., Chiang, J.H.y., Lafuente, A.L.: Sok: lending pools in
  decentralized finance. In: International Conference on Financial Cryptography
  and Data Security. pp. 553--578. Springer (2021)

\bibitem{black1973pricing}
Black, F., Scholes, M.: The pricing of options and corporate liabilities.
  Journal of political economy  \textbf{81}(3),  637--654 (1973)

\bibitem{bonneau2015sok}
Bonneau, J., Miller, A., Clark, J., Narayanan, A., Kroll, J.A., Felten, E.W.:
  {Sok: Research perspectives and challenges for bitcoin and cryptocurrencies}.
  In: Security and Privacy (SP), 2015 IEEE Symposium on. pp. 104--121. IEEE
  (2015)

\bibitem{daian2020flash}
Daian, P., Goldfeder, S., Kell, T., Li, Y., Zhao, X., Bentov, I., Breidenbach,
  L., Juels, A.: Flash boys 2.0: Frontrunning in decentralized exchanges, miner
  extractable value, and consensus instability. In: 2020 IEEE Symposium on
  Security and Privacy (SP). pp. 910--927. IEEE (2020)

\bibitem{darlin2020optimal}
Darlin, M., Papadis, N., Tassiulas, L.: Optimal bidding strategy for maker
  auctions. arXiv preprint arXiv:2009.07086  (2020)

\bibitem{eskandari2021sok}
Eskandari, S., Salehi, M., Gu, W.C., Clark, J.: Sok: Oracles from the ground
  truth to market manipulation. In: Proceedings of the 3rd ACM Conference on
  Advances in Financial Technologies. pp. 127--141 (2021)

\bibitem{hull2003options}
Hull, J.C.: Options futures and other derivatives. Pearson Education India
  (2003)

\bibitem{kao2020analysis}
Kao, H.T., Chitra, T., Chiang, R., Morrow, J.: An analysis of the market risk
  to participants in the compound protocol. In: Third International Symposium
  on Foundations and Applications of Blockchains (2020)

\bibitem{klages2020stablecoins}
Klages-Mundt, A., Harz, D., Gudgeon, L., Liu, J.Y., Minca, A.: Stablecoins 2.0:
  Economic foundations and risk-based models. In: Proceedings of the 2nd ACM
  Conference on Advances in Financial Technologies. pp. 59--79 (2020)

\bibitem{klages2019stability}
Klages-Mundt, A., Minca, A.: (in) stability for the blockchain: Deleveraging
  spirals and stablecoin attacks. arXiv preprint arXiv:1906.02152  (2019)

\bibitem{klages2022while}
Klages-Mundt, A., Minca, A.: While stability lasts: A stochastic model of
  noncustodial stablecoins. Mathematical Finance  (2022)

\bibitem{bitcoin}
Nakamoto, S.: {Bitcoin: A peer-to-peer electronic cash system}  (2008)

\bibitem{perez2021liquidations}
Perez, D., Werner, S.M., Xu, J., Livshits, B.: Liquidations: Defi on a
  knife-edge. In: International Conference on Financial Cryptography and Data
  Security. pp. 457--476. Springer (2021)

\bibitem{qin2021cefi}
Qin, K., Zhou, L., Afonin, Y., Lazzaretti, L., Gervais, A.: Cefi vs.
  defi--comparing centralized to decentralized finance. arXiv preprint
  arXiv:2106.08157  (2021)

\bibitem{qin2021empirical}
Qin, K., Zhou, L., Gamito, P., Jovanovic, P., Gervais, A.: An empirical study
  of defi liquidations: Incentives, risks, and instabilities. In: Proceedings
  of the 21st ACM Internet Measurement Conference. pp. 336--350 (2021)

\bibitem{qin2022quantifying}
Qin, K., Zhou, L., Gervais, A.: Quantifying blockchain extractable value: How
  dark is the forest? In: 2022 IEEE Symposium on Security and Privacy (SP). pp.
  198--214. IEEE (2022)

\bibitem{qin2021attacking}
Qin, K., Zhou, L., Livshits, B., Gervais, A.: Attacking the defi ecosystem with
  flash loans for fun and profit. In: International Conference on Financial
  Cryptography and Data Security. pp. 3--32. Springer (2021)

\bibitem{shreve2005stochastic}
Shreve, S.: Stochastic calculus for finance I: the binomial asset pricing
  model. Springer Science \& Business Media (2005)

\bibitem{stoll1969relationship}
Stoll, H.R.: The relationship between put and call option prices. The Journal
  of Finance  \textbf{24}(5),  801--824 (1969)

\bibitem{wang2022speculative}
Wang, Z., Qin, K., Minh, D.V., Gervais, A.: Speculative multipliers on defi:
  Quantifying on-chain leverage risks. In: Financial Cryptography and Data
  Security: 26th International Conference, FC 2022, Grenada, May 2--6, 2022,
  Revised Selected Papers. pp. 38--56. Springer (2022)

\bibitem{whelan2001economic}
Whelan, J., Msefer, K., Chung, C.V.: Economic supply \& demand. MIT (2001)

\bibitem{wood2014ethereum}
Wood, G., et~al.: Ethereum: A secure decentralised generalised transaction
  ledger. Ethereum project yellow paper  \textbf{151}(2014),  1--32 (2014)

\bibitem{zhou2021high}
Zhou, L., Qin, K., Torres, C.F., Le, D.V., Gervais, A.: High-frequency trading
  on decentralized on-chain exchanges. In: 2021 IEEE Symposium on Security and
  Privacy (SP). pp. 428--445. IEEE (2021)

\bibitem{zhou2022sok}
Zhou, L., Xiong, X., Ernstberger, J., Chaliasos, S., Wang, Z., Wang, Y., Qin,
  K., Wattenhofer, R., Song, D., Gervais, A.: Sok: Decentralized finance (defi)
  attacks. arXiv preprint arXiv:2208.13035  (2022)

\end{thebibliography}

\appendix




\section{Black-Scholes Model}
\label{app:bs-model}
We apply the Black-Scholes model~\cite{black1973pricing} to price call options under optimal assumptions, such as the non-existence of dividend  payouts. 
The option premium is calculated for European call options on a per-share basis. The payoff for $\sellerShort$ introduced in Figure~\ref{figure:payoffCurve} is trivial to grasp but it does not yield any insights on the pricing of the option.
    With the BS model for a European call option determines the option price as
    \begin{equation}
        c = \spotExchangeRate e^{-\foreignInterestRate \cdot \terminationTime} N(d_1) - \strikePriceShort e^{-\domesticInterestRate \cdot \terminationTime} N(d_2)
    \end{equation}
    where
    \begin{equation}
        d_1 = \frac{\ln(\spotExchangeRate\/\strikePriceShort)+(\domesticInterestRate-\foreignInterestRate+\volatilityAsset^2\/2) \cdot \terminationTime}{\volatilityAsset\cdot \sqrt{\terminationTime}}
    \end{equation}
    and 
    \begin{equation}
        d_2 = d_1 - \volatilityAsset\cdot \sqrt{\terminationTime}.
    \end{equation}
    $\spotExchangeRate$ is the spot exchange rate, $\foreignInterestRate$ is the foreign interest rate, $\domesticInterestRate$ is the domestic interest rate and $\volatilityAsset$ is the volatility of the underlying asset.
    For a detailed introduction to the Black-Scholes pricing model for European call options, we refer the interested reader to~\cite{hull2003options}.
    
We remark that the B-S model does not take into account the decrease in risk and lowered average payoff due to termination by $\sellerShort$. We defer a more precise pricing model for \financialPrimitives that to future work.

\clearpage
\section{Tables}\label{app:tables}

\begin{table}[]
    \centering
    \caption{Accumulative collateral restraint by \protocol over a time-frame of~\liquidationTimeFrame.}
    \begin{tabular}{cc|cccccccccc}
    \toprule
    \protocol Premium Factor $\lambda$ &&& $1\%$ && $2\%$ && $5\%$ && $10\%$ && $20\%$ \\\midrule
     Accumulative Collateral Restraint (USD) &&& \CollateralRestraintOne && \CollateralRestraintTwo && \CollateralRestraintFive && \CollateralRestraintTen && \CollateralRestraintTwenty\\
    \bottomrule
    \end{tabular}
    \label{tab:accumulative_collateral_restraint}
\end{table}

\begin{table}[]
\centering
\caption{Payoffs for \protocol supporters at maturity assuming that borrowers would not rescue. We present the probability that a supporter \textit{(i)} exercises the call option and profits, \textit{(ii)}  exercises the call option but loses, \textit{(iii)} defaults. We also simulate the average profit for supporters. our simulations are based on the real market prices.}
\resizebox{\columnwidth}{!}{%
\begin{tabular}{c|ccccc|c}
\toprule
           $\lambda$           & $1\%$ & $2\%$ & $5\%$ & $10\%$ & $20\%$                  &     $\Delta T$                      \\ \midrule
\multicolumn{1}{c|}{$+$} & $87.46\%$    &   $87.46\%$  &   $87.46\%$  &   $87.46\%$   & \multicolumn{1}{c|}{$87.46\%$} & \multirow{4}{*}{1hour}    \\
\multicolumn{1}{c|}{$-$} & $0.29\%$    &  $0.58\%$   &  $1.41\%$   &  $2.47\%$    & \multicolumn{1}{c|}{$4.14\%$} &                           \\
\multicolumn{1}{c|}{$\#$} & $12.25\%$   &  $11.96\%$   &  $11.13\%$   &  $10.08\%$    & \multicolumn{1}{c|}{$8.41\%$} &                           \\
\multicolumn{1}{c|}{$\$$} &  $125.51$K$\pm1.52$M   &  $125.51$K$\pm1.52$M    &  $125.50$K$\pm1.52$M    &  $125.49$K$\pm1.52$M     & \multicolumn{1}{c|}{$125.48$K$\pm1.52$M} &                           \\\midrule
\multicolumn{1}{c|}{$+$} & $87.19\%$    &  $87.19\%$   &  $87.19\%$   &   $87.19\%$   & \multicolumn{1}{c|}{$87.19\%$} & \multirow{4}{*}{6 hours}  \\
\multicolumn{1}{c|}{$-$} &  $0.30\%$   &   $0.60\%$  &   $1.50\%$  &  $2.68\%$    & \multicolumn{1}{c|}{$4.44\%$} &                           \\
\multicolumn{1}{c|}{$\#$} &  $12.51\%$   &   $12.21\%$  &  $11.31\%$   &  $10.13\%$    & \multicolumn{1}{c|}{$8.37\%$} &                           \\
\multicolumn{1}{c|}{$\$$} &  $154.01$K$\pm2.19$M   &  $154.01$K$\pm2.19$M   &  $154.00$K$\pm2.19$M   &  $154.00$K$\pm2.19$M    & \multicolumn{1}{c|}{$154.98$K$\pm2.19$M} &                           \\\midrule
\multicolumn{1}{c|}{$+$} & $85.95\%$    &  $85.95\%$   &   $85.95\%$  &  $85.95\%$    & \multicolumn{1}{c|}{$85.95\%$} & \multirow{4}{*}{24 hours} \\
\multicolumn{1}{c|}{$-$} &  $0.56\%$   &  $1.03\%$   &  $2.16\%$   &   $3.62\%$   & \multicolumn{1}{c|}{$5.59\%$} &                           \\
\multicolumn{1}{c|}{$\#$} &  $13.48\%$   &  $13.02\%$   &  $11.89\%$   &  $10.42\%$    & \multicolumn{1}{c|}{$8.45\%$} &                           \\
\multicolumn{1}{c|}{$\$$} & $144.42$K$\pm1.83$M    &  $144.40$K$\pm1.83$M   &   $144.36$K$\pm1.83$M  &   $144.32$K$\pm1.83$M   & \multicolumn{1}{c|}{$144.29$K$\pm1.83$M} &                           \\\bottomrule
\end{tabular}%
}
\begin{tabular}{lllll}
$+$ exercise and profit &\qquad\qquad& $-$ exercise but lose &\qquad\qquad& $\#$ default\\
\multicolumn{5}{l}{$\$$ average profit for supporters in USD (mean$\pm$std)}
\end{tabular}
\label{tab:payoffs}
\end{table}

\end{document}